\documentclass[useAMS,a4paper,usenatbib]{mn2e}
\usepackage[dvips]{graphicx}
\usepackage{epsfig}
\usepackage{amsmath}
\usepackage{subfigure} 
\usepackage{hyperref}
\usepackage{amssymb}
\title[Signatures of mergers in disc kinematics] {Signatures of
  minor mergers in Milky Way-like disc kinematics: Ringing revisited}

\author[F.~A. G\'omez et al.]{Facundo A. G\'omez$^{1,2}$\thanks{Email:fgomez@pa.msu.edu},  
Ivan Minchev$^{3}$, 
\'Alvaro Villalobos$^{4}$, \newauthor 
Brian W. O'Shea$^{1,2,5}$
\& Mary E. K. Williams$^{3}$
\\
$^{1}$ Department of Physics and Astronomy, Michigan State University, East Lansing, MI 48824, USA\\
$^{2}$ Institute for Cyber-Enabled Research, Michigan State University, East Lansing, MI 48824, USA\\
$^{3}$ AIP, An der Sterwarte 16, 14482 Potsdam, Germany\\
$^{4}$ INAF-Osservatorio Astronomico di Trieste, Via Tiepolo 11,I-34143 Trieste, Italy\\
$^{5}$ Lyman Briggs College, Michigan State University, East Lansing, MI 48825, USA\\
}

\begin{document}

\date{}

\pagerange{\pageref{firstpage}--\pageref{lastpage}} \pubyear{}

\maketitle

\label{firstpage}

\begin{abstract}

  By means of $N$-body simulations we study the response of a galactic
  disc to  a minor merger  event.  We find  that non-self-gravitating,
  spiral-like  features are  induced in  the thick  disc.  As  we have
  shown in a previous work, this ``ringing'' also leaves an imprint in
  velocity space (the $u$-$v$ plane) in small spatial regions, such as
  the  solar neighbourhood.   As  the disc  relaxes  after the  event,
  clumps in  the $u$-$v$  plane get closer  with time, allowing  us to
  estimate  the  time  of  impact.   In  addition  to  confirming  the
  possibility  of  this  diagnostic,  here  we show  that  in  a  more
  realistic scenario,  the in-fall  trajectory of the  perturber gives
  rise to an azimuthal dependence  of the structure in phase-space. We
  also find that the space  defined by the energy and angular momentum
  of stars  is a better choice  than velocity space,  as clumps remain
  visible even in large local volumes.  This makes their observational
  detection much  easier since one need  not be restricted  to a small
  spatial volume. We  show that information about the  time of impact,
  the  mass of  the perturber,  and its  trajectory is  stored  in the
  kinematics of disc stars.

\end{abstract}


\section{Introduction}
The  present  day  spatial  distribution, motion,  ages  and  chemical
abundances of  stars contain important information  about how galaxies
form and evolve \citep{FBH}.   Substructure in these distributions has
been      observed,     especially      in      the     Milky      Way
\citep[e.g.][]{ibata94,hwzz99,new02,ibata03,yanny03,nhf,belu06,h06,else09,klement09,will09,will11},
but            also            in            other            galaxies
\citep[e.g.][]{ibata01a,chap,Mart08,Mart09,McCo09}.  This substructure
can arise  from a variety of different  physical mechanisms.  Examples
are the  disruption of open clusters  \citep[][]{eggen}, the dynamical
effects induced by non-axysimmetric components  of a galaxy, such as a
bar      or      self-gravitating      spiral     arms      \citep[see
e.g.][]{walter,fux01,min08,ant09,min10,quill11},    or   debris   from
satellite  galaxies that  were  accreted by  their  more massive  host
\citep[e.g.][]{hw,hz00,bj05,font,sales08,md11}.     Another   possible
mechanism, introduced by \citet[][hereafter M09]{min09}, is related to
the  response of  a  galactic disc  to  the tidal  interaction with  a
satellite  galaxy  as  it   merges.   Thanks  to  dynamical  friction,
relatively massive satellites can reach  the centre of the disc before
being completely disrupted by  the tidal field \citep{bt}.  M09 showed
that the sudden energy kick imparted by the gravitational potential of
the  satellite, as  it crosses  the plane  of the  disc,  can strongly
perturb the velocity field of disc stars located in local volumes such
as the  Solar Neighbourhood.  These  perturbations can be  observed in
the $u$-$v$ plane as arc-like  features travelling in the direction of
positive $v$, where $u$ and $v$ are the radial and tangential velocity
components.  This  mechanism, known as ``ringing'',  could explain the
presence of high velocity streams observed in the Solar Neighbourhood,
such as the Arcturus  stream \citep{will09}.  Furthermore, as shown by
\citet{quill09}, minor  mergers can also  induce radial mixing  in the
outer disc, thus partially  explaining the lack of correlation between
age  and   metallicity  observed  in   the  Milky  Way's   thick  disc
\citep[see][and references therein]{alvphd}.

An important limitation in the M09 analysis is the extent of the local
volume that can be explored to search for signatures of ringing in the
$u$-$v$  plane.   As   the  depth  of  the  volume   is  increased,  a
progressively larger number of  density waves is included. This causes
the  waves  to  interfere,  wiping  out  the  substructure  previously
observed  in this  projection of  phase-space.  This  not  only places
constraints on  the size  of stellar samples  that can be  explored to
search for signatures of ringing  in the Solar Neighbourhood, but also
makes  studying  this  mechanism  in  fully  self-consistent  $N$-body
simulations unfeasible.   Note that to resolve local  volumes as small
as, e.g., 100 pc, an  extremely large particle resolution is required.
Thus, in M09 only  relatively simple high-resolution 2D simulations of
massless test-particles were performed to explore this problem.

In this paper we show that the space of energy and angular momentum is
much  more suitable  for identifying  the signatures  of ringing  in a
galactic disc.   In this projection  of phase-space much  larger local
volumes can  be explored  without erasing the  substructure associated
with  ringing (Section~\ref{sec:test_part_sim}).   This  allows us  to
analyse  a set  of  fully self-consistent  $N$-body  simulations of  a
merger   between  a   relatively  massive   satellite  galaxy   and  a
pre-existing thin disc (Section~\ref{sec:nbody}).  We briefly describe
our       simulations       in       Section~\ref{sec:sims}.        In
Section~\ref{sec:conclu}  we  summarise our  results  and discuss  its
application  to currently  available as  well as  forthcoming  full 6D
phase-space samples of thick disc stars in the Solar vicinity.

\section{Simulations}
\label{sec:sims}
Two types of numerical simulations are studied in this work. First, we
analyse  a two  dimensional  massless test-particle  simulation of  an
unrelaxed stellar galactic disc, as  described by M09.  In this model,
a flat rotational velocity profile  is considered, with a value of 220
km s$^{-1}$ at 7.8 kpc from  the galactic centre.  The disc follows an
exponential  density profile  with a  scale-length of  3  kpc. Initial
velocities for the particles  are drawn from Gaussian distributions in
the $u$ and $v$ directions.   In particular, the standard deviation in
the  radial direction is  set at  $\sigma_{u} =  50$ km  s$^{-1}$.  To
emulate  a  population that  is  unrelaxed,  velocities are  purposely
chosen  so that  particles are  unevenly distributed  in  their radial
oscillation.  A total of $7 \times 10^{6}$ test-particles were used to
represent the disc.  For more details, we refer the reader to M09.

Second, we analyse a set of fully self-consistent $N$-body simulations
of the merger  between a satellite galaxy and  a pre-existing galactic
thin  disc.   Most  of  these  simulations  were  first  presented  by
\citet[][hereafter, VH08]{vh08}  to study the formation  of a galactic
thick  disc  due to  the  heating induced  by  the  merger event.   In
particular, the simulations considered in this work are those referred
as ``$z=1"$ in VH08, which were  set up to emulate a merger event that
could have given  rise to the Milky Way's thick  disc.  The models for
the host  and satellite galaxies consist of  two fully self-consistent
components:  a   dark  matter  (DM)  halo  following   a  NFW  profile
\citep{nfw},  a  stellar  disc for  the  host,  and  a bulge  for  the
satellite  system.  The  stellar disc  follows an  exponential density
profile, and the satellite's  bulge a Hernquist profile \citep{hernq}.
The properties  of the Milky  Way-like host and the  satellite systems
were  scaled to those  expected at  $z =  1$ following  \citet{mo}. As
described by \citet{brooks},  in this picture gas in  a halo conserves
its specific angular momentum (equal to that of the dark matter) as it
cools  to form  a centrifugally  supported  disc that  grows from  the
inside out.   In the simplest model,  in which the  density profile of
the galaxy is modelled as  a singular isothermal sphere, the radius of
the resulting disc  is proportional to the parent  halo virial radius.
Thus, in a $\Lambda$-CDM cosmology, this would imply a decrease in the
disc size of a factor of ~1.7 out to $z = 1$. Later studies have shown
that the evolution of the  disc size - stellar mass relation predicted
by  this model  is considerably  stronger than  what  observations are
suggesting  \citep[see e.g.][]{somer}. Note  however that  the results
presented in  this work are  not strongly dependent on  the particular
value chosen for the host disc's scale-length.

\begin{table}
\begin{minipage}{90mm} \centering
  \caption{Initial  properties  of   the  three  $N$-body  simulations
    analysed in this work.}
\label{table:model}
\begin{tabular}{@{}llllr} \hline Host's DM halo & & & & \\
\hline 
Virial mass, $M^{\rm DM}_{\rm host}$ & $5.07 \times 10^{11}$ & & & $[M_{\odot}]$ \\ 
Virial radius & $122.22$ & & & [kpc] \\
Concentration & 6.56  & & & \\
\hline
Host's stellar disc & & & & \\
\hline
Mass, $M^{\rm stellar}_{\rm host}$ & $1.2 \times 10^{10}$ & & & [$M_{\odot}$] \\
Scale length, $R_{\rm d}$ & 1.65 & & & [kpc] \\
Scale height, ${\rm z}_{0}$ & 0.165 & & & [kpc] \\
\hline
Satellite's DM halo & $f =20\%$ & $15\%$ & $10\%$ &\\
\hline
Virial mass, $M^{\rm DM}_{\rm sat}$ & $1.01$ & $0.75$ & $0.5$ & $10^{11} [M_{\odot}]$ \\ 
Virial radius & $71.35$ & $64.65$ & $56.63$ & [kpc] \\
Concentration & 8.09 & 8.41 & 8.86 & \\
\hline  
Satellite's stellar bulge & & & & \\
\hline
Mass, $M^{\rm stellar}_{\rm sat}$ & $2.4$ & $1.8$ & $1.2$ & $10^{9}[M_{\odot}$] \\
Scale radius, $r_{\rm b}$ & 0.709 & 0.654 & 0.583 & [kpc] \\
\end{tabular}
\end{minipage}
\end{table}

Both disc  and bulge are  represented with $1 \times  10^{5}$ $N$-body
particles.  To study  the impact that mergers of  different total mass
ratios have on the phase-space  structure of the disc, three different
values of  $f = M^{\rm  tot}_{\rm host} /M^{\rm  tot}_{\rm sat}$
were   explored.    These  are   $f=20\%$,   $15\%$,   and  $10\%$   .
Table~\ref{table:model}  summarises  the   values  of  the  parameters
involved in  these models.  Note  that $f$ also sets  $M^{\rm DM}_{\rm
  host} /M^{\rm  DM}_{\rm sat}$ and  $M^{\rm stellar}_{\rm host}
/M^{\rm stellar}_{\rm sat}$.

The  following is  a  brief summary  of  the evolution  of the  merger
simulations analysed in this study.  We refer the interested reader to
VH08  for a  complete description.   In  all cases  the satellite  was
released from a distance of 84 kpc ($\approx 50 ~ R_{\rm d}$) from the
galactic  centre  in the  most  likely  infalling  orbit according  to
previous studies  on distributions of orbital  parameters of infalling
substructure  \citep[e.g.][]{ben05}.   Here $R_{\rm  d}$  is the  host
disc's scale-length.   In such rather eccentric  orbit ($e=0.86$), the
satellite  rapidly reaches  the central  region of  the  galactic halo
where it is more effectively  affected by dynamical friction and tidal
disruption.   This  causes  the  satellite  to  lose  orbital  angular
momentum, which progressively shrinks  its infalling orbit.  In such a
way,  for the  simulation with  $f =  20\%$, $15\%$,  and  $10\%$, the
satellite takes  roughly $t_{\rm merger}$ \footnote{ In  this work the
  merger  time, $t_{\rm  merger}$, is  defined  as the  time when  the
  satellite  runs out  of orbital  angular momentum.  For the  kind of
  orbits considered in this work, the oscillations between the centres
  of  mass of  both systems  ceases approximately  at the  same time.}
$\approx 2$, $2.5$ and $4$ Gyr to reach and settle in the disk centre,
respectively.   In   particular,  for  the   $f=20\%$,  the  satellite
experiences $\sim 8$ pericentric passages, all of them at less than 20
kpc from the  disk centre.  As the satellite decays  in its orbits, it
induces the formation  of tidal arms in the  disc, which carry angular
momentum from the disk centre outwards.  Once the satellite is brought
on  to the  plane  of the  disc  by dynamical  friction, it  transfers
kinetic energy  from its  orbits to the  disc stars,  increasing their
vertical motions and causing a  visible thickening.  By the end of the
simulations the  morphological, structural and  kinematical properties
of  the thickened  disc and  the disrupted  satellite settle  down and
become stable, which  takes place roughly ~2 Gyr  after the satellites
have  reached  the  disk  centre.   The  final  scale-heights  of  the
resulting thick disks  are $\approx 5 ~ {\rm  z}_{0}$ for $f=20\%$ and
$\approx 4 ~ {\rm z}_{0}$ for  $f=10\%$ (see Figure 14 of VH08).  Here
${\rm z}_{0}$ is the scale-height of the pre-existing thin disc.

\section{Test-particle simulations of an unrelaxed disc}
\label{sec:test_part_sim}

\begin{figure*}
\hspace{-0.05cm}
\includegraphics[width=174.5mm,clip]{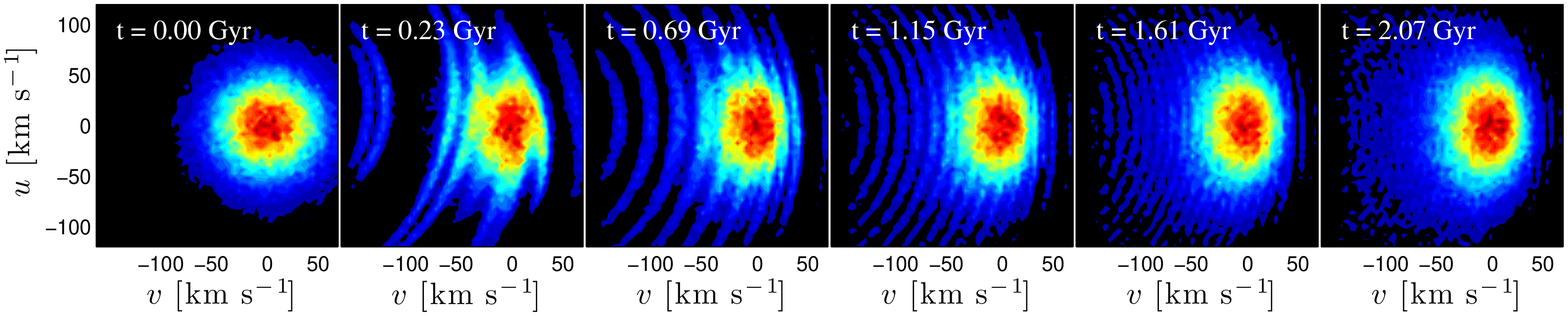}
\\
\includegraphics[width=175mm,clip]{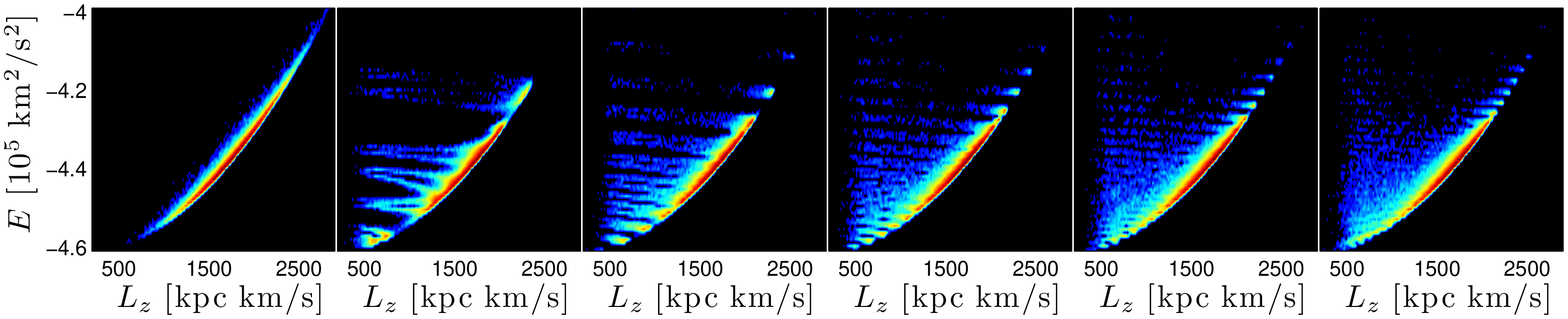}
\caption{Distribution  of particles located  within a  ring of  160 pc
  width, centred  at 7.8 kpc  from the galactic centre,  obtained from
  our test-particles  simulation of an unrelaxed  disc.  The different
  colours (contours) indicate different number of particles. The first
  column  of panels  shows the  distribution of  particles  before the
  ``merger'' event.   Top panels:  Time development of  the $u$  - $v$
  plane distribution of particles.  Note that, as explained in M09, as
  time passes  by the  number of arc-like  features increases  and get
  closer together.  Bottom panels: Time development of the $E$-$L_{z}$
  distribution of  particles.  The substructure  present in the  $u$ -
  $v$ can  be now  observed as group  of particles  with approximately
  constant energy covering a wide range of angular momentum.}
\label{fig:model_uv_el}
\end{figure*}

In this  section we briefly review  the main results  presented in M09
and study how the observed substructure in the velocity space of local
volumes is  distributed in  $E$-$L_{z}$ space.  In  the top  panels of
Figure~\ref{fig:model_uv_el}  we  show, at  six  different times,  the
distribution in the {\it u-v} plane of particles located within a ring
of 160 pc width centred at 7.8 kpc from the disc centre. Note that the
simulation is axysimmetric.  Thus, the  rings can be regarded as local
volumes. As  a consequence of the  sudden energy kick  imparted on the
initially relaxed  disc, a radial  oscillation is excited on  the disc
particles. The  radial frequency of  these orbits will depend  on both
their orbital energies and  angular momenta (see M09). Thus, particles
with different energies will  have different radial frequencies. These
differences in frequencies induce phase-wrapping in velocity space: at
a given  time after the perturbation,  in a given volume  of space, we
will  observe groups  of  particles that  have  completed a  different
number of  radial oscillations. Each one  of these groups  will have a
characteristic frequency  or energy. These  groups can be  observed in
the u-v plane as distinct arc-like features. Note that, as time passes
by and the disk relaxes, the arcs get closer together and their number
increases.  The  bottom panels of  Fig.~\ref{fig:model_uv_el} show the
same distribution  of particles now projected  onto $E$-$L_{z}$ space.
Particles  that previously were  part of  a given  arc in  the $u$-$v$
plane  are  now  distributed  along well-defined  features  of  nearly
constant energy. Due to the wide range covered in the {\it v} velocity
component, each  arc is stretched  along the $L_{z}$  direction.  Note
that,  within  a volume  as  small as  the  one  considered here,  the
gravitational  potential, $\Phi(r)$,  can be  regarded as  a constant.
Thus, for groups of particles with approximately constant energy,
\begin{equation}
\nonumber
v^{2} + u^{2} = 2 (E - \Phi(r)) \approx cst, 
\end{equation}
which explains the arc-like shape  of the features observed in the {\it
  u-v}  plane.   

As pointed  out  by  M09,  due to  the  differential
rotation of the galactic disc, the separation between arcs in the {\it
  u-v}  plane depends  on the  galactocentric distance  of  the volume
under study.  Increasing the size  of this volume, and thus sampling a
larger range  of galactocentric radii, causes the  waves to interfere,
wiping out most of the  observed substructure in velocity space.  This
places strong constraints on the size  of the sample of stars that can
be explored to identify signatures of ringing in the solar vicinity if
we use velocity space.

\begin{figure}
\hspace{-0.11cm}
\includegraphics[width=85.3mm,clip]{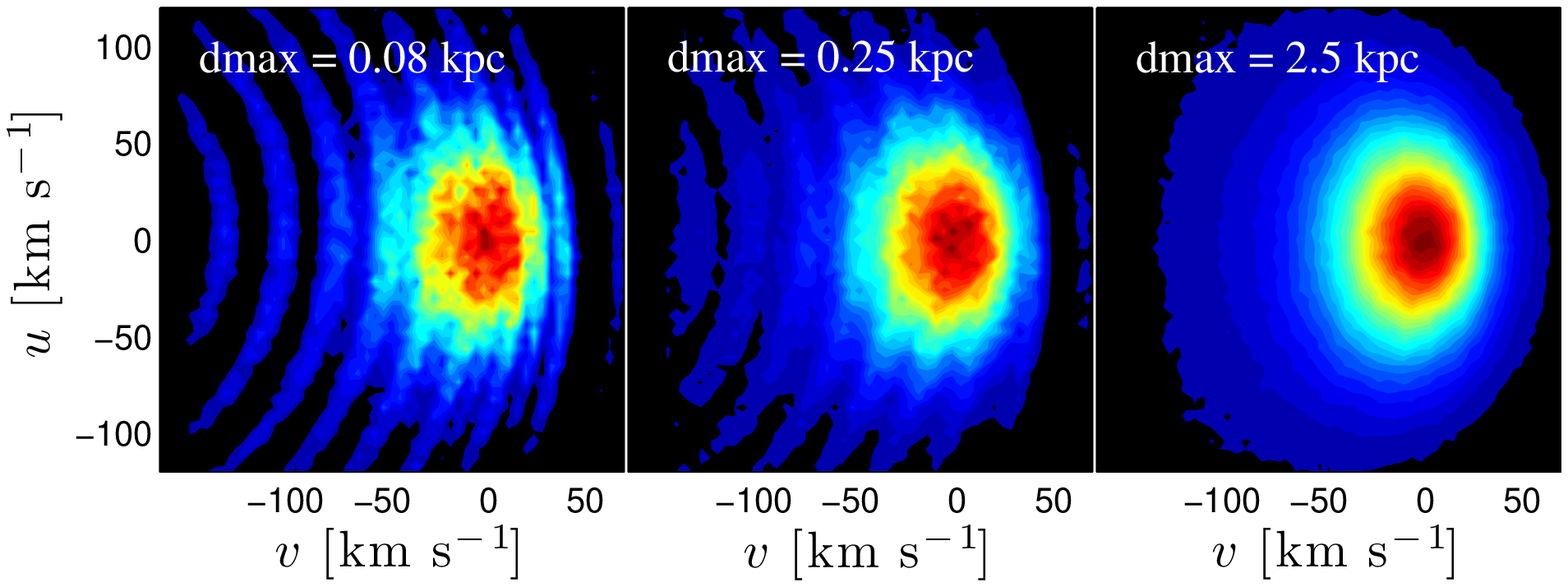}
\\
\includegraphics[width=85mm,clip]{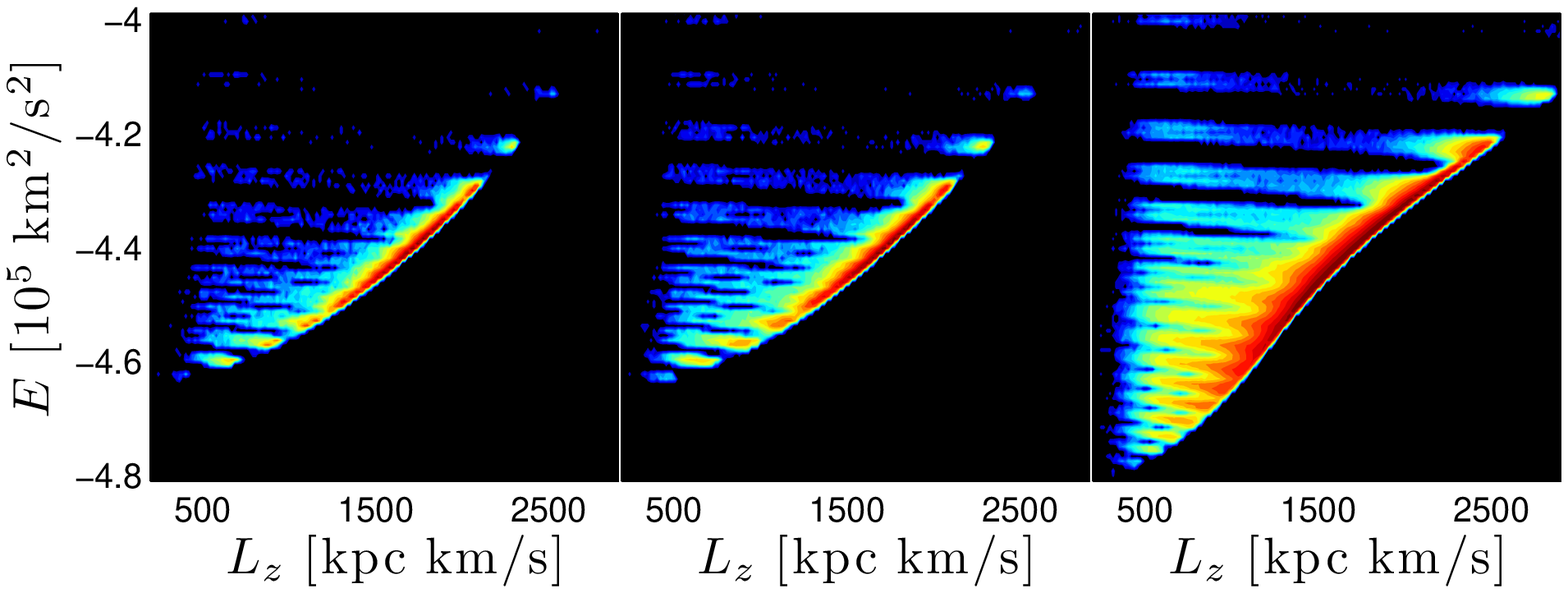}
\caption{Distribution  of particles  within rings  centred at  7.8 kpc
  from   the  galactic  centre,   obtained  from   our  test-particles
  simulation of  an unrelaxed disc.  The  different colours (contours)
  indicate different  number of  particles. Each column  is associated
  with a  ring of a different  width, as indicated in  the top panels.
  All distribution are  drawn from the same snapshot,  $t = 0.69$ Gyr.
  As the width of the ring is increased, the arcs in the $u$-$v$ plane
  (top  panels)  become  less  clear.   At $d_{\rm  max}  =  2.5$  kpc
  substructure has been completely  wiped out.  Note, however, that in
  $E$-$L_{z}$  space   substructure  can  be   clearly  observed.   As
  expected,  the  number of  features  in  this  space increases  with
  increasing $d_{\rm max}$.  Note  as well that the separation between
  energy waves decreases as a function of energy (see text).}
\label{fig:model_dist}
\end{figure}

In Figure~\ref{fig:model_dist}  we compare the effect  of volume depth
in the $u$-$v$ and $E$-$L_{z}$ planes for the same simulation and time
step.   The top  row we  show the  velocity distribution  of particles
located inside  rings with different widths.   The mean galactocentric
distance  $r_{\rm mean}$  of  each ring  is  7.8 kpc  and the  maximum
distance sampled  from $r_{\rm mean}$  by these rings is  specified in
each panel  as d$_{\rm max}$.  Note  that, by including  in our sample
particles that are only at 250  pc away from $r_{\rm mean}$, the waves
start to  interfere and the  arcs in {\it  u-v} to overlap. In  a ring
with d$_{\rm max} = 2.5$  kpc all the substructure has been completely
erased.   The bottom  panel of  Figure~\ref{fig:model_dist}  shows the
distribution of the same sets  of particles in $E$-$L_{z}$ space. When
very small  local volumes are  considered, i.e.  d$_{\rm max}  = 0.08$
kpc,   both  spaces   are  equivalent   in  terms   of   the  observed
substructure. However, it is very  interesting to see that even in the
ring with largest d$_{\rm max}  = 2.5$ kpc substructure can be clearly
observed in $E$-$L_{z}$ space.  By including particles at larger radii
we  include density waves  at different  energy levels.   In addition,
each wave is  better defined since particles that  were in a different
phase of their oscillation, and thus outside the arc, are now included
in the sample.  Note that, as expected, the separation between density
waves  decreases as a  function of  energy, i.e.,  with galactocentric
radius.

The results presented  in this section are important  for two reasons.
First, by sampling  relatively large local volumes, it  is possible to
study  the properties  of  ringing in  fully self-consistent  $N$-body
simulations, where the  particle resolution is much lower  than in the
previous experiment.  Second, future astrometric missions such as {\it
  Gaia}  \citep{gaia} will  provide very  accurate measurements  of 6D
phase-space coordinates for stars located at distances as large as 2.5
kpc from the position  of the Sun \citep[see e.g.][]{gea10}.  Although
the  $u$-$v$  plane and  the  $E$-$L_{z}$  space  are equivalent  when
searching for  signatures of  ringing in local  spheres with  radii as
small as 80  pc, the fraction of thick disc stars  in this volume will
be considerably smaller than in a  sphere of 2.5 kpc radius. Note that
it is  in the thick disc where  we would expect to  find signatures of
ringing, since these stars are less affected by perturbations from the
Galactic bar,  spiral structure or giant molecular  clouds.  Thus, the
$E$-$L_{\rm z}$ space will allow us  to explore much larger samples of thick
disc stars from, e.g., a future {\it Gaia} catalog.

\section{Fully self-consistent $N$-body simulations}
\label{sec:nbody}

\begin{figure}
\hspace{0.001cm}
\includegraphics[width=42.5mm,clip]{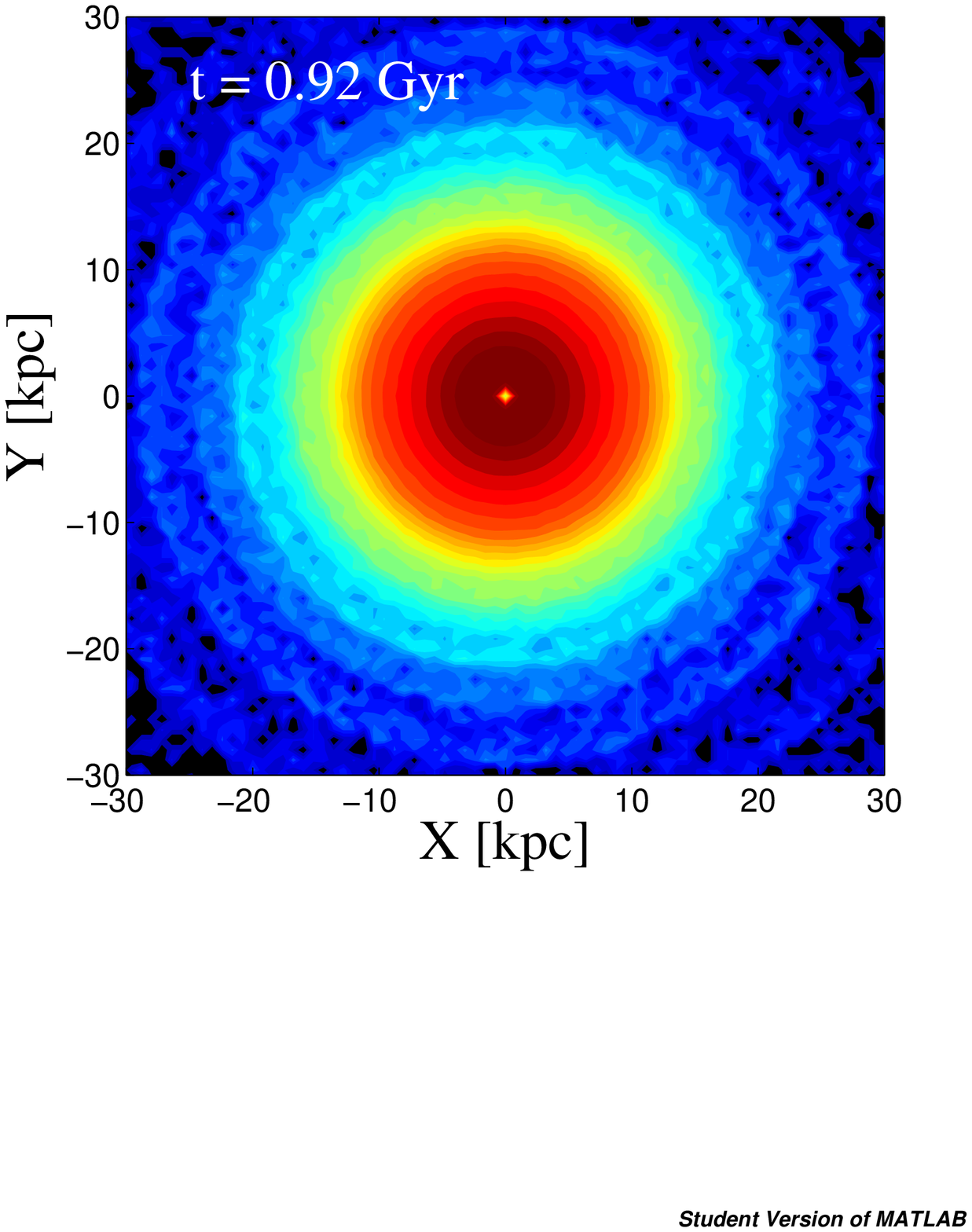}
\hspace{-0.07cm}
\includegraphics[width=38mm,clip]{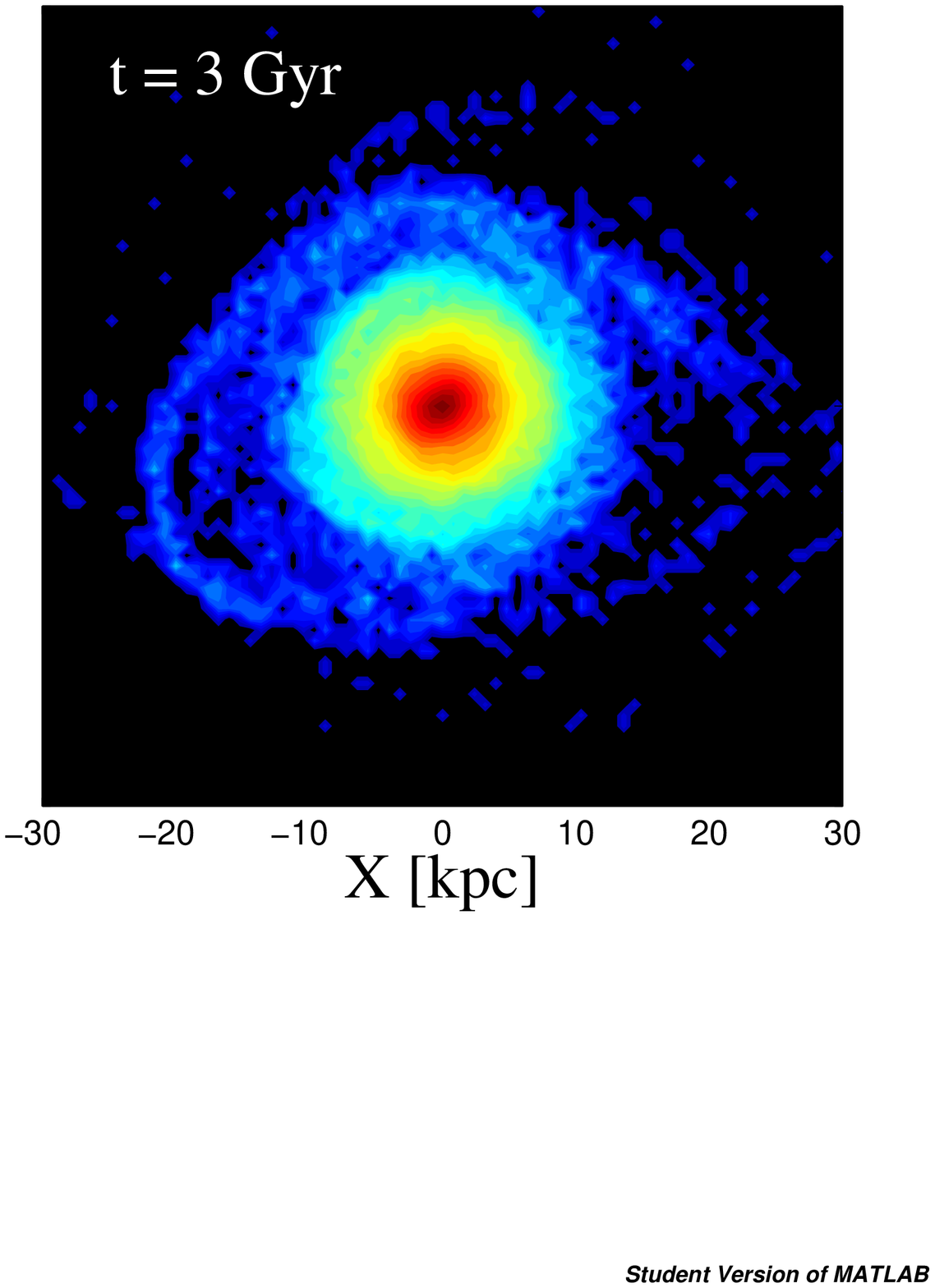}
\caption{Left panel: Distribution of particles  in the X - Y plane, as
  obtained from  our test-particles  simulation of an  unrelaxed disc,
  $\approx  1$  Gyr after  the  energy  kick.   The different  colours
  (contours) indicate different number of particle.  Note the multiple
  annular  features,  associated  with  density waves.   Right  panel:
  Distribution of  particles in the  X - Y  plane 1 Gyr  after $t_{\rm
    merger}$ (t = 3 Gyr),  as obtained from our $N$-body simulation of
  a merger event with a mass ratio of $f = 20\%$.  As a consequence of
  the non-axysimmetrical  energy kick imparted by the  satellite as it
  crosses the plane of the disc, spiral features (rather than annular)
  are formed.}
\label{fig:xy_space}
\end{figure}

\begin{figure}
\includegraphics[width=80mm,clip]{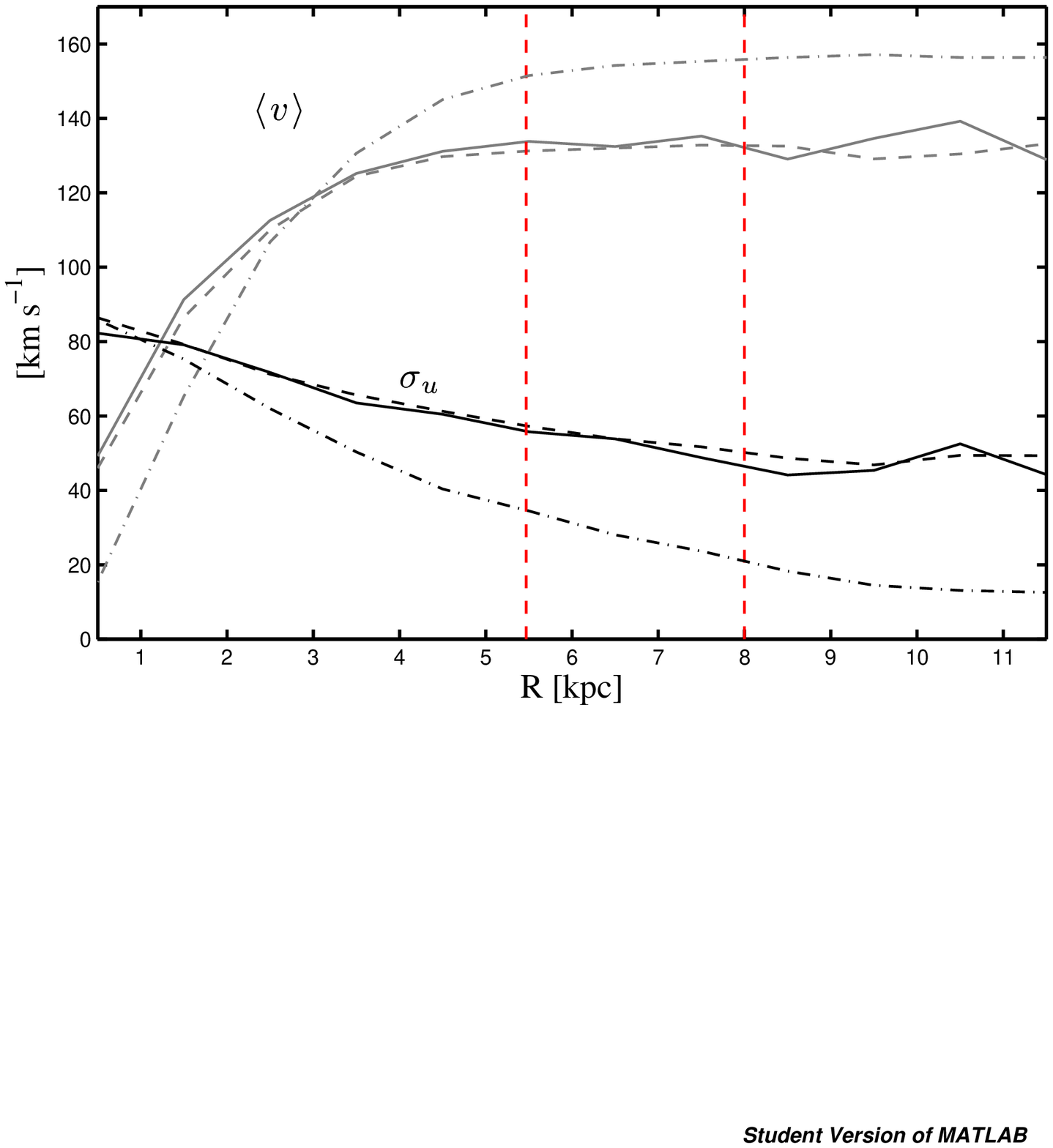}
\caption{Radial  velocity  dispersion   and  mean  azimuthal  velocity
  profiles  at  three  different  times, from  our  $f=20\%$  $N$-body
  simulation. The profiles have  been computed within concentric rings
  of 1 kpc width for particles  with $|{\rm Z}| < 1.5$ kpc.  The solid
  and dashed  lines show  the profiles computed  at $t=2.1$  and $4.5$
  Gyr, which  corresponds to 0.1  and 2.5 Gyr after  $t_{\rm merger}$,
  respectively.  The  dotted dashed  lines represent both  profiles at
  $t=0$ Gyr. Note  that the later  were obtained after evolving  the host
  for  3  Gyr in  isolation.   The  red  vertical lines  indicate  the
  positions of the spheres studied in Section~\ref{sec:dens_el}}
\label{fig:tm_vel}
\end{figure}

We  now explore  fully self-consistent  $N$-body simulations  of minor
merger  events.  As  described by  VH08, the  spatial  and kinematical
structures of the pre-existing thin  disc are strongly affected by the
merger event.  As the satellite decays in its orbit, it passes through
the plane of  the disc several times. At  each passage, disc particles
are  impulsively  heated   by  the  satellite's  potential  \citep[see
e.g.,][VH08]{qh93}.  One  manifestations of this  heating mechanism is
the   formation  of   transient,   non-self-gravitating  spiral   arms
\citep[see e.g.][]{tutu,oh}.  The  right panel of Figure~\ref{fig:xy_space}
shows a  contour plot of  the distribution of  particles in the X  - Y
plane of our $f = 20\%$  simulation, 1 Gyr after $t_{\rm merger}$.  It
is very interesting to compare this distribution with the one obtained
from the  test-particles simulation,  shown in the  left panel  of the
same figure.  Since the energy  kick imparted on the test-particles is
axisymmetric  annular features  (left panel)  rather than  spiral arms
(right panel), can be observed.  It is important to emphasise that the
spiral  features   observed  in  our  $N$-body   simulations  are  not
self-gravitating, as they quickly  disappear with time.  This can also
be inferred from the large values of the radial velocity dispersion of
the resulting  thick disc  shown in Figure~\ref{fig:tm_vel}.   In this
figure we show  the radial velocity dispersion profile  of the disc at
three different times. The  radial velocity dispersion is considerably
larger  at all radii  already 0.1  Gyr after  the satellite  galaxy is
fully merged.   This profile does not evolve  significantly during the
next 2.9 Gyr of simulation.  Within  5 to 10 kpc from the disc centre,
the radial  velocity dispersion takes  a value of  $\sigma_{u} \approx
50$  km s$^{-1}$,  which  is consistent  with  the value  used in  the
test-particles simulation.  As  expected, the asymmetric drift induced
in this way causes a decrease in the rotational velocity as a function
of  radius  after  the  merger.   Note that,  although  its  magnitude
decreased  by about  $20\%$  at $R  \geq  4$ kpc,  a  flat profile  is
preserved.   We will now  analyse solar-neighbourhood-like  volumes in
the  resulting thick  disc, to  search  for signatures  of ringing  in
energy and angular momentum space.

\subsection{Density waves in $E$-$L_{z}$ space}
\label{sec:dens_el}

Since the particle resolution of  these simulations is much lower than
that of our test-particles simulation,  it is not feasible to look for
signatures of density  waves in volumes as small  as the ones analysed
in  Section~\ref{sec:test_part_sim}.  In  what follows,  we  focus our
analysis on  samples of  particles located within  spheres of  2.5 kpc
radius.   The  simulations analysed  in  this  work  were designed  to
emulate the result  of a minor merger experienced  by a Milky Way-like
galaxy at  $z=1$. Therefore,  the scales of  our final thick  disc are
relatively smaller  compared to those  observed in the Milky  Way.  As
explained by  \citet{vh09}, it is not straightforward  to decide which
radii  corresponds to  a solar  neighbourhood-like  volume.  Following
their work, we place our sphere at two different locations, namely 5.5
kpc ($\sim 2.4  R_{\rm td}$) and 8 kpc ($\sim  3.5 R_{\rm td}$), where
$R_{\rm td}  = 2.28$ kpc  is the final  scale length of  the simulated
thick disc.  This  also allows us to study  the expected dependance of
the  number  and distribution  of  density  waves with  galactocentric
radius.

\begin{figure}
\hspace{0.001cm}
\includegraphics[width=41.5mm,clip]{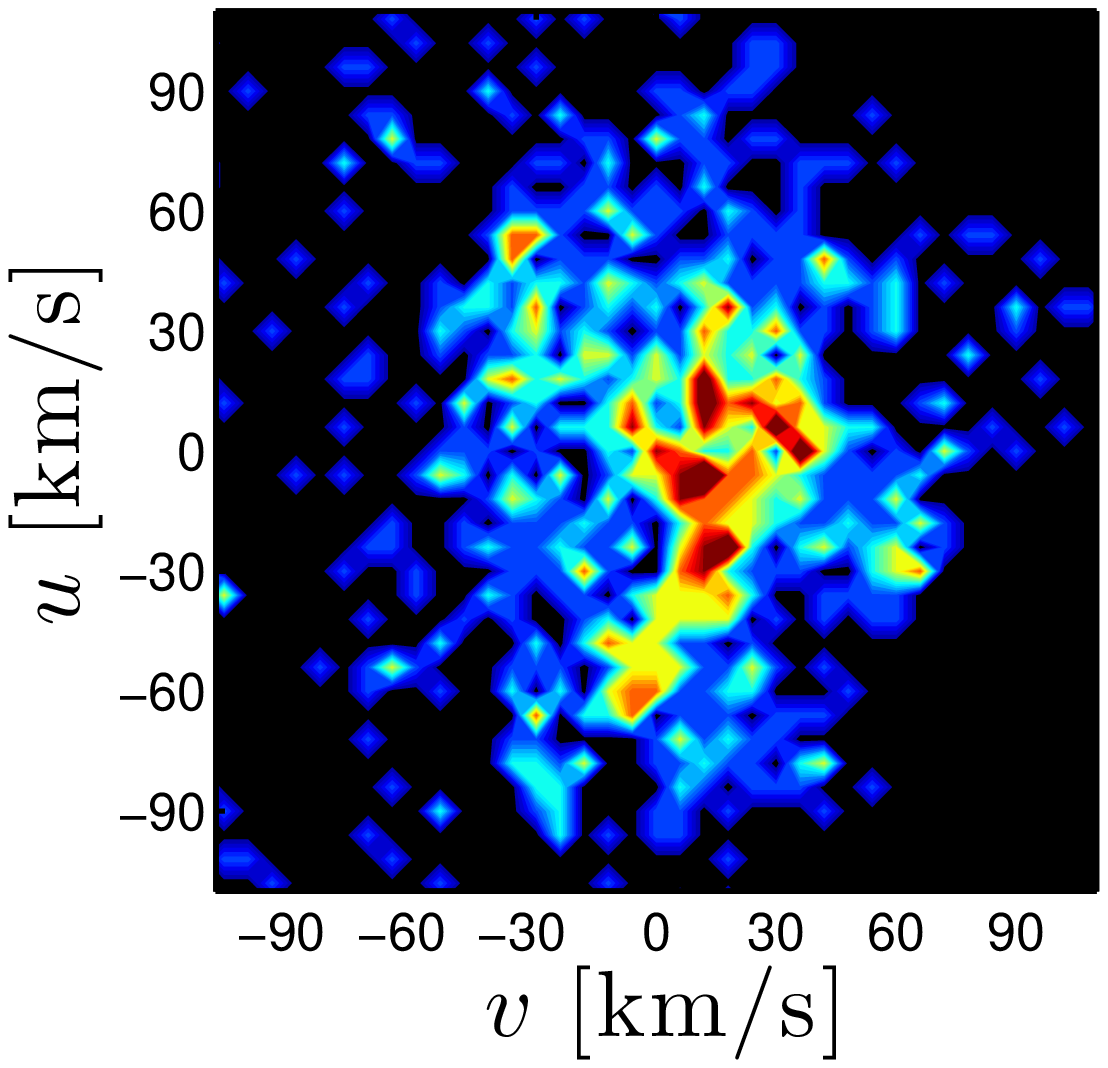}
\includegraphics[width=41.5mm,clip]{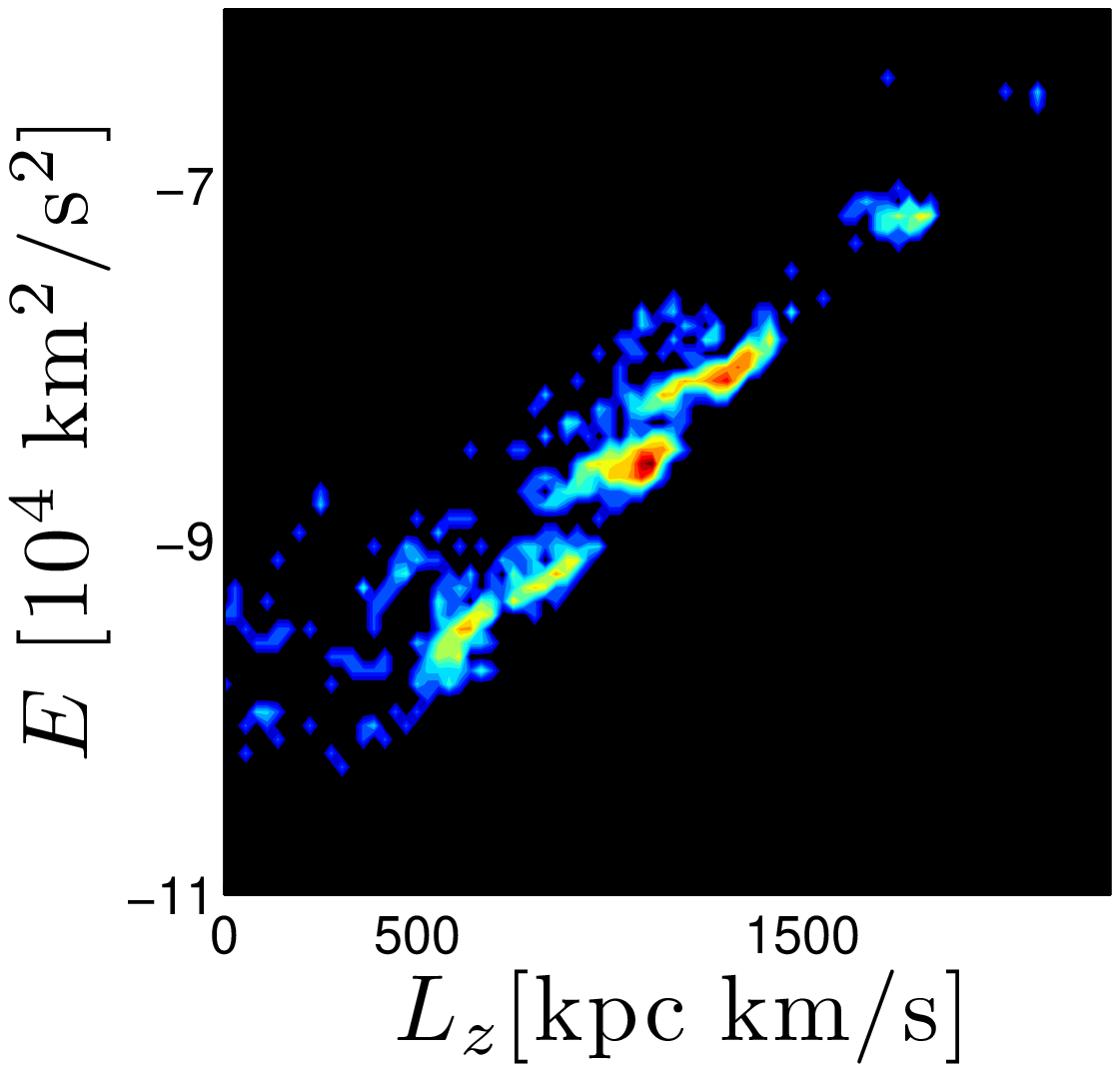}
\caption{Distribution  of disc particles  inside a  sphere of  2.5 kpc
  radius located at 8 kpc  from the centre, obtained from our $N$-body
  simulation with  a mass ratio of  $f = 20\%$, 0.7  Gyr after $t_{\rm
    merger}$. The  left panel shows  the distribution of  particles in
  velocity  space   whereas  the   right  panel  corresponds   to  the
  $E$-$L_{z}$ projection. Note that, unlike in velocity space, a large
  number of substructures can be observe in $E$-$L_{z}$ space.}
\label{fig:comp_uv_elz}
\end{figure}

\begin{figure*}
\hspace{-0.05cm}
\includegraphics[width=175.9mm,clip]{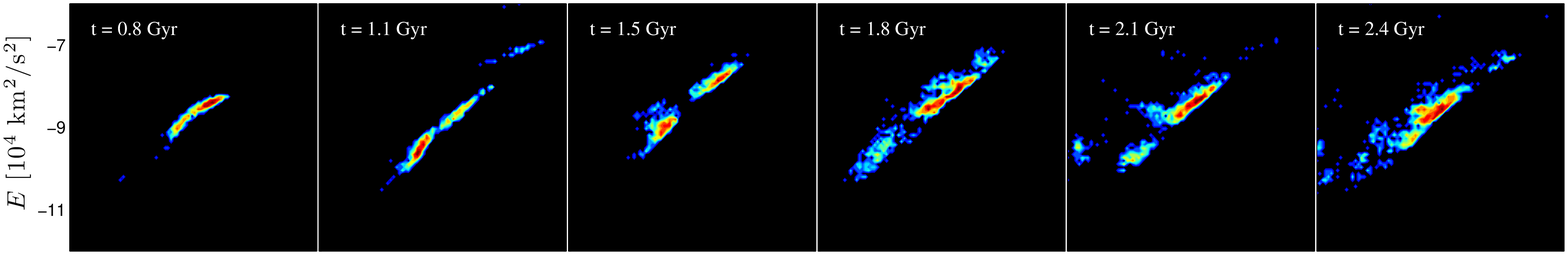}
\\
\includegraphics[width=176.5mm,clip]{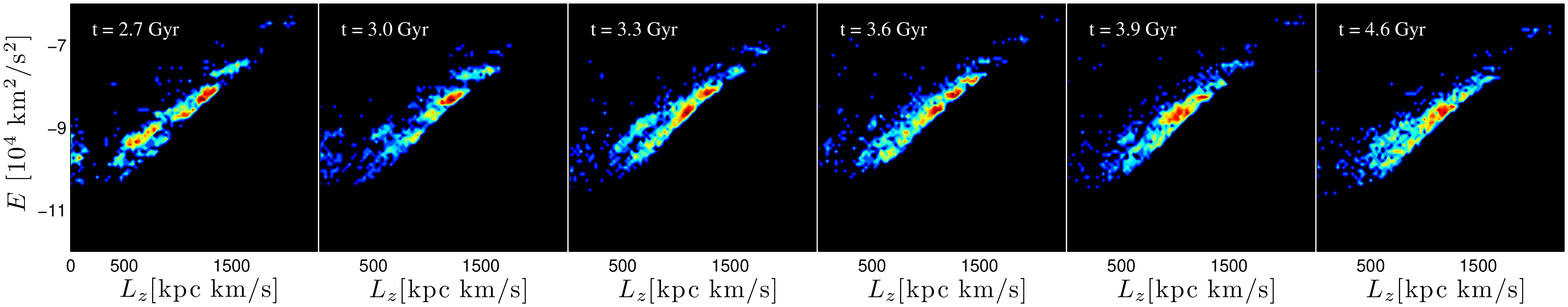}
\caption{Time development of the  $E$-$L_{z}$ space for a distribution
  of particles  located within a  sphere of 2.5  kpc centred at  8 kpc
  from the galactic centre, obtained from our $N$-body simulation with
  a mass ratio of $f =  20\%$.  The sphere is rotating with an angular
  velocity set by the velocity of the LSR. Note that, as expected, the
  number  of lumps  associated with  density waves  increases  as time
  passes  by.  Substructure  can be  clearly observed  for a  total of
  $\sim 4$ Gyr. In this simulation the merger concludes at $t_{\rm
    merger} \approx 2$ Gyr.}
\label{fig:tot_mass_8kpc}
\end{figure*}

\begin{figure*}
\includegraphics[width=176mm,clip]{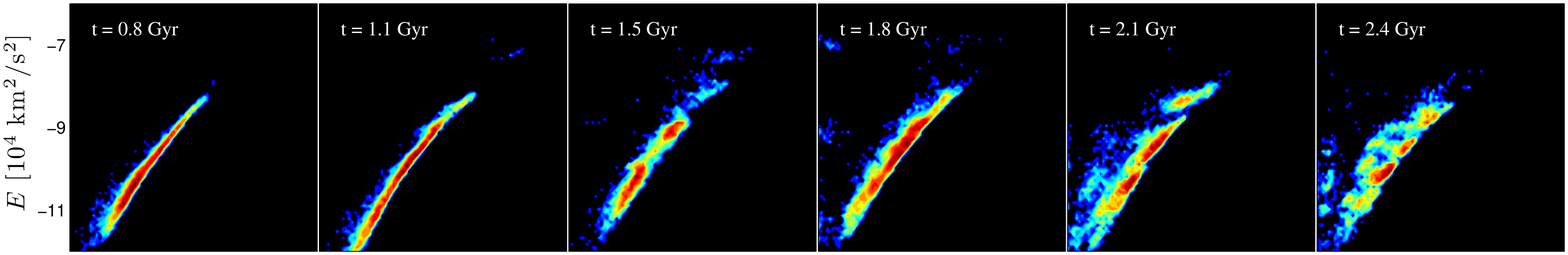}
\\
\hspace{-0.1cm}
\includegraphics[width=176.3mm,clip]{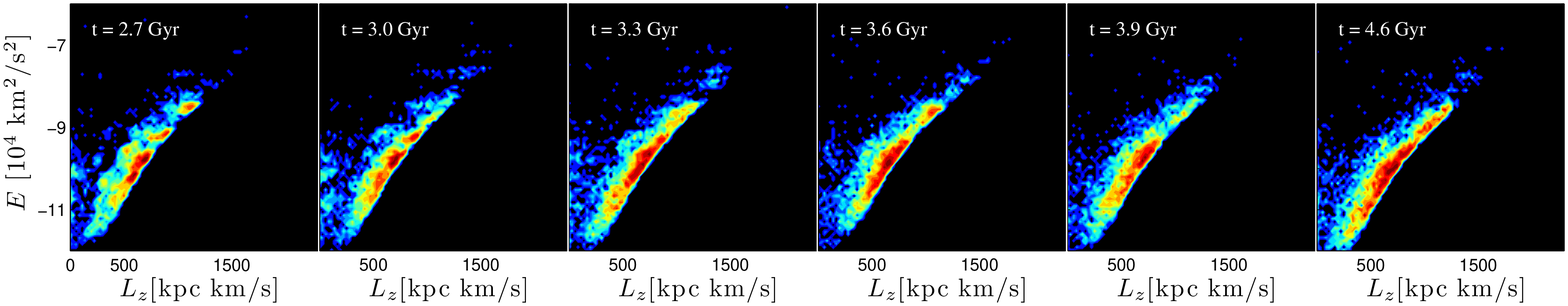}
\caption{As  in Figure~\ref{fig:tot_mass_8kpc}  for  particles located
  within a sphere  centred at 5.47 kpc from  the galactic centre. Note
  that mixing occurs more rapidly than in our sphere centred at 8 kpc,
  reflecting the dependance of the mixing time-scale on galactocentric
  radius. In  this sphere, substructure  can be clearly observe  for a
  total of to $\sim 2.8$  Gyr. In this simulation the merger concludes
  at $t_{\rm merger} \approx 2$ Gyr.}
\label{fig:tot_mass_5.47kpc}
\end{figure*}

The left panel  of Figure~\ref{fig:comp_uv_elz} shows the distribution
in  the $u-v$  plane of  particles that  are located  inside  a sphere
centred at (8,0,0) kpc from the galactic centre, 0.7 Gyr after $t_{\rm
  merger}$\footnote{  This particular  time and  location  were chosen
  arbitrarily for simplicity.   Dependencies with azimuthal angles are
  explored  in   Section~\ref{sec:azi_dep}}.   This  distribution  was
obtained  from  our $f=20\%$  simulation.   Note  that  only 865  disc
particles  are found  within this  sphere.   Due to  the large  volume
sampled,  the  absence of  arc-like  features  in  this projection  of
phase-space  is not  surprising.  The  same distribution  of particles
projected in  $E$-$L_{z}$ space  is shown on  the right panel  of this
figure.  Interestingly, particles  are now distributed in well-defined
clumps  of approximately  constant  energy.  Each  clump represents  a
different       density      wave      crossing       the      sphere.
Figure~\ref{fig:tot_mass_8kpc}  shows  the   time  evolution  of  this
distribution.  Note that our solar neighbohood-like volume is rotating
with an angular frequency set by the velocity of the local standard of
rest  (LSR),  obtained  after  correcting the  final  mean  rotational
velocity profile  for axysimmetric drift \citep{nma}.   Over time, the
number of clumps increases and  they come closer together, as expected
from Section~\ref{sec:test_part_sim} (see also M09).  Substructure can
be clearly observed already at $t \approx 0.8$ Gyr and can be followed
until $t  \approx 4.7$ Gyr, spanning  a total of about  4 Gyr.  Later,
although  still  present, the  signatures  of  the  density waves  are
difficult to observe even in  $E$-$L_{z}$ space.  Recall that for this
simulation, $t_{\rm merger} \approx 2$ Gyr.  Thus, substructure can be
observe in the $E$-$L_{z}$ plane for about 2.7 Gyr after the satellite
has   completely  ``sunk''  to   the  centre   of  the   host's  disc.
Figure~\ref{fig:tot_mass_5.47kpc}  shows, as a  function of  time, the
$E$-$L_{z}$ distribution  of the  particles located inside  the sphere
centred  at 5.47 kpc  from the  galactic centre.   Again, a  wealth of
substructure  can  be  observed.    Comparison  of  this  figure  with
Figure~\ref{fig:tot_mass_8kpc} reveals the presence of a larger number
of clumps at any given time after $t_{\rm merger}$.  This reflects the
dependance of  the phase-mixing time-scale  on galactocentric distance
(see  M09).   Note that  in  this  inner  sphere the  perturbation  is
triggered after $\sim 1.1$ Gyr,  a time that corresponds to the second
pericentric passage (and  thus, closer to the galactic  centre) of the
satellite galaxy.  Substructure can be clearly observed for a total of
$\approx 2.8$ Gyr, and for $\approx 1.9$ Gyr after $t_{\rm merger}$.

We have also analysed a  self-consistent simulation of a galactic disk
evolved  in isolation.   The  disc  is first  relaxed  within a  rigid
version  of the  halo potential  for about  1 Gyr.   Once the  disc is
relaxed the rigid  halo is replaced by its  N-body counterpart and the
system is evolved fully self-consistently for 4 Gyr (See VH08, Section
A3 for a detailed description of  this procedure). Even after 4 Gyr of
evolution,  no noticeable  substructure can  be observed  in  the E-Lz
distribution  of  particles  located within  solar  neighbourhood-like
volumes.

\subsection{Azimuthal dependance}

\label{sec:azi_dep}

\begin{figure*}
\includegraphics[width=170mm,clip]{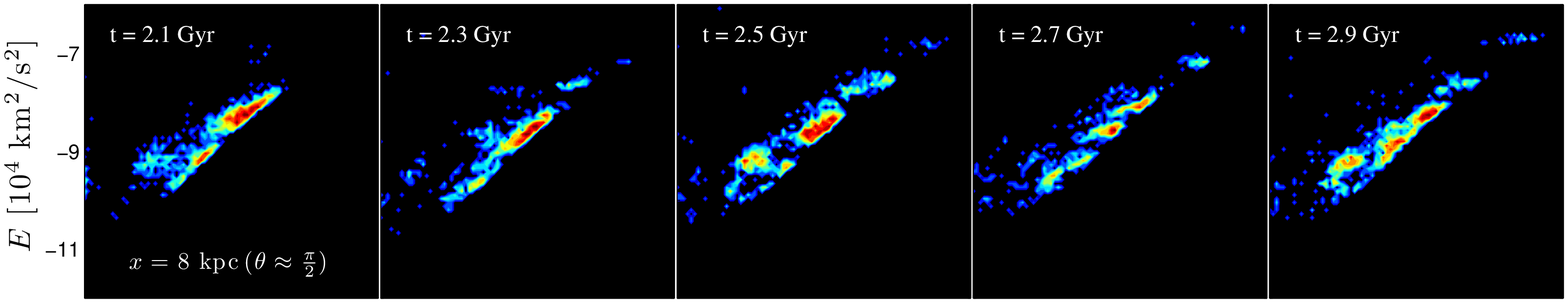}
\\
\hspace{-0.13cm}
\includegraphics[width=170.4mm,clip]{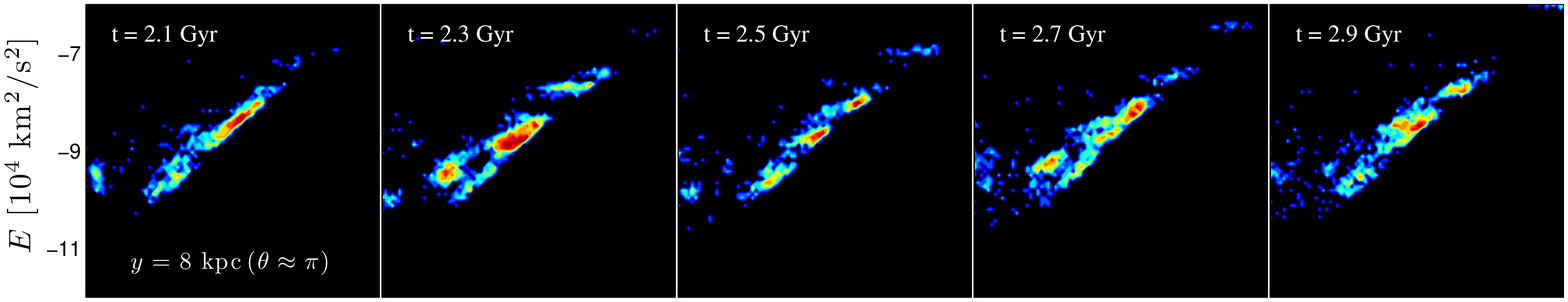}
\\
\includegraphics[width=170.1mm,clip]{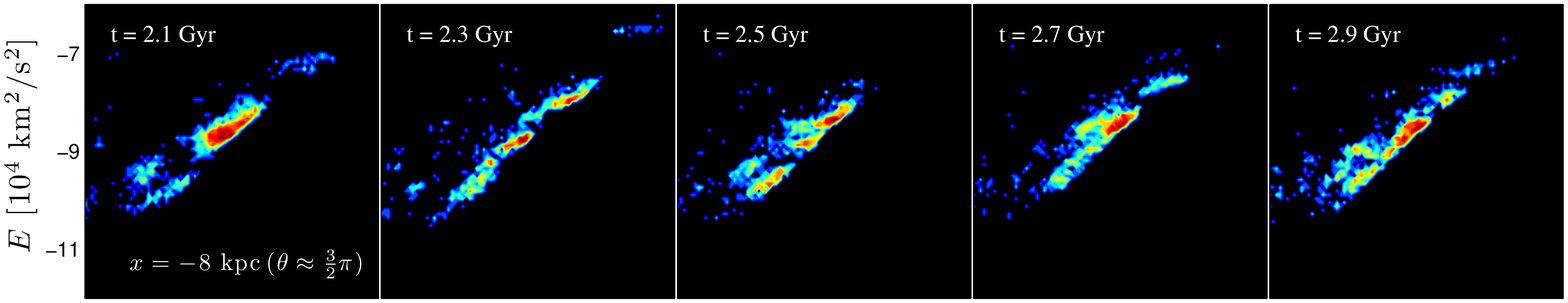}
\\
\hspace{-0.09cm}
\includegraphics[width=169.9mm,clip]{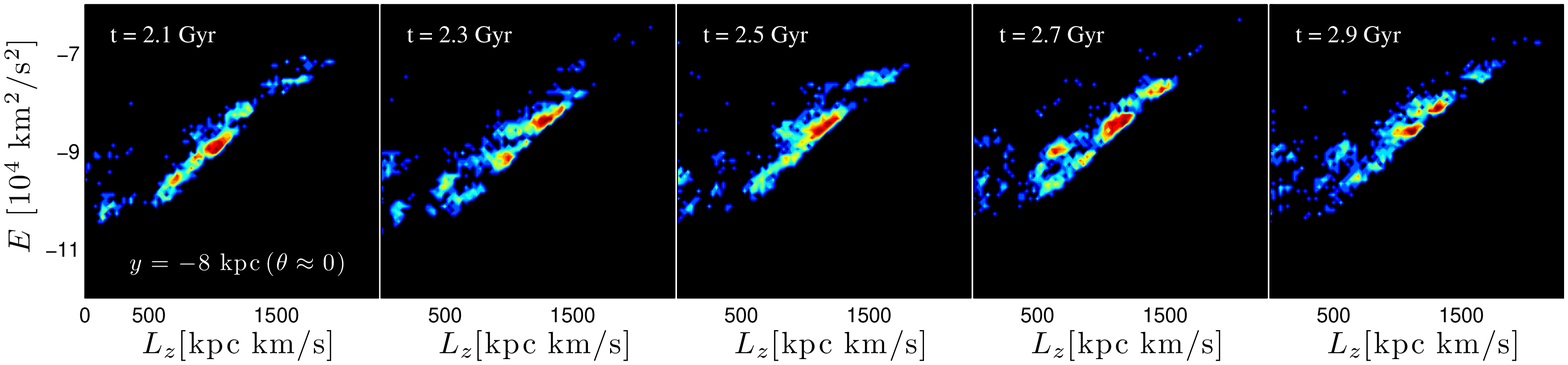}
\caption{Azimuthal  dependance   on  the  distribution   of  lumps  in
  $E$-$L_{z}$ space. Each  row of panel shows the  time development of
  the  distribution  of  disc  particles inside  a  different  sphere,
  obtained  from our $N$-body  simulation with  a mass  ratio of  $f =
  20\%$.  The location  of the  spheres are  indicated on  the panel's
  bottom   left  corner   of  each   panel.   The   variable  $\theta$
  approximately indicates  the angle with respect to  the direction of
  the  first close  passage  of  the satellite  through  plane of  the
  disc. The spheres  are kept fixed in their  initial locations.  Note
  that, along diagonals starting from bottom left to top right panels,
  the  distribution of particles  approximately repeats  itself.  This
  indicates both, a phase difference on the density waves at different
  azimuthal  angles   and  a  relation  between   angular  and  radial
  frequencies of disc particles.}
\label{fig:azimuth}
\end{figure*}

As  opposed   to  the  axisymmetric   energy  kick  imparted   to  our
test-particle simulation, the merging  satellite galaxy is expected to
impulsively heat the disc  in a preferential direction associated with
its orbit.  Therefore,  we would expect to observe  a phase difference
in    the   distribution   of    substructure   observed    in   solar
neighbourhood-like volumes located  at different azimuthal angles.  To
explore  this, we follow  in our  $f=20\%$ simulation  the $E$-$L_{z}$
distribution of  the particles enclosed within  four different spheres
for $\sim  1$ Gyr.   The spheres are  located at $(\pm8,0,0)$  kpc and
$(0,\pm 8,0)$ kpc and hereafter we  will refer to them as $x=\pm8$ kpc
and  $y=\pm8$ kpc.  For this  particular analysis  we keep  the sphere
fixed  at   their  initial  locations.   The  results   are  shown  in
Figure~\ref{fig:azimuth}. Each row of  panels shows the time evolution
of this distribution  on a different sphere. The  panels are labelled,
in the leftmost column, according to the position of the corresponding
sphere.   At  any given  time,  the  distribution  of substructure  in
$E$-$L_{z}$  space  is different  from  sphere  to  sphere.  A  visual
inspection of the  first column of panels indicates  a larger fraction
of  substructure on  the  $y=-8$  kpc sphere,  0.1  Gyr after  $t_{\rm
  merger}$.  Note  that this direction corresponds  with the direction
of the first {\it close} passage of the satellite through the galactic
disc, as shown by Figures 1 and 2 in VH08.  It is also noticeable that
the distribution of particles  repeats itself along diagonals starting
from bottom  left to  top right panels.   This implies that  the phase
difference  of the density  wave can  be well  described as  an spiral
pattern,  travelling  approximately  an  angle of  $\dfrac{3}  {2}\pi$
radians in 0.2 Gyr, in agreement with the spatial distribution of disc
particles observed in Figure~\ref{fig:xy_space}.  To examine this, let
us estimate the averaged angular period of a particle located at 8 kpc
from the galactic centre.  From Figure~\ref{fig:tm_vel}, at $r=8$ kpc,
$\langle v \rangle \approx 130$ km s$^{-1}$. Thus,
\begin{equation}
\nonumber
T_{\phi} = \dfrac{2\pi r}{\langle v \rangle} \approx 0.38 {\rm ~Gyr},
\end{equation}
implying that  in 0.2 Gyr, particles have  travelled approximately an
angle $\pi$ through the disc.  As a consequence, the ratio between the
angular and  the radial  frequencies\footnote{Note that most  of these
  particles  have  very  low  eccentricities and  thus  $\kappa  \approx
  \Omega_{r}$, where  $\kappa$ represents the  epyciclic frequency} of
the particles located inside our sphere is expected to be
\begin{equation}
\nonumber
\dfrac{\Omega_{\phi}}{\Omega_{r}} \approx \dfrac{2}{3}.
\end{equation}

\begin{figure}
\includegraphics[width=80mm,clip]{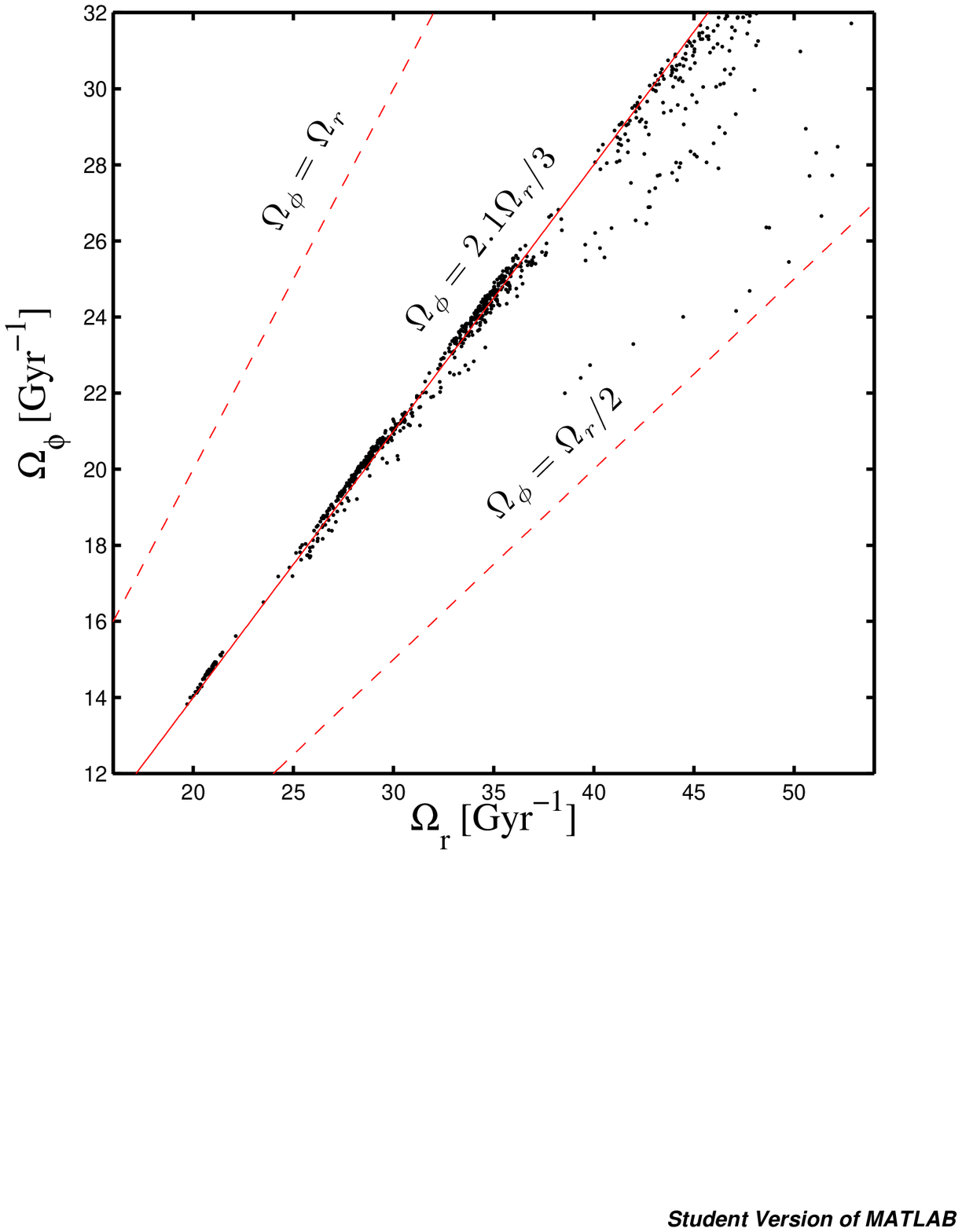}
\caption{Distribution in  frequency space  of disc particles  inside a
  sphere of 2.5 kpc radius located  at 8 kpc from the galactic centre,
  obtained  from our $N$-body  simulation with  a mass  ratio of  $f =
  20\%$,  0.7 Gyr  after $t_{\rm merger}$.  Note  that substructure  can be
  easily identify as well in  this space. The solid line indicates the
  ratio  between frequencies  of  disc particles,  whereas the  dashed
  lines denote the limits of this space.}
\label{fig:freqs}
\end{figure}

Figure~\ref{fig:freqs} shows  the distribution of  orbital frequencies
of the particles located inside the $x = +8$ kpc sphere, 0.7 Gyr after
$t_{\rm  merger}$.   The frequencies  were  computed  as described  in
\citet[][hereafter GH10]{gh10}  and we refer the reader  to their work
for  a  detailed description  of  their  estimation method  \citep[see
also][]{daniel}.   Note  that  particles  tend  to lie  on  the  curve
$\Omega_{\phi}  = 2.1  \Omega_{r} /  3$,  in good  agreement with  our
previous  discussion.  It  is  very interesting  to  observe that,  in
frequency space, particles are  also distributed in well-defined lumps
representing the different density waves crossing the volume.
   
\subsection{Dependance on satellite's mass}

It is well  known that the amount of disc heating  induced by a merger
event  strongly depends on  the properties  of the  merging satellite,
such    as   its    concentration   and    total    mass   \citep[see,
e.g.,][VH08]{qh93,ben04,dm11}.  Although, since $z  = 1$, mergers with mass
ratios as large as $20\%$ are  likely to occur in galaxies such as the
Milky Way,  the likeliness of  these events increases  with decreasing
mass  ratios  \citep[see  e.g.][]{lh,kaza09,cooper}. It  is  therefore
interesting  to study  whether perturbations  induced by  less massive
minor mergers  can leave imprints  on the phase-space  distribution of
galactic  disc stars.   In this  section we  investigate  minor merger
events     with    mass     ratios     of    $f     =    15\%$     and
$10\%$. Table~\ref{table:model} summarises the main properties of both
satellites.  For  less massive satellites than  those considered here,
dynamical friction is expected to  be less efficient.  Thus, this kind
of satellites are  unlikely to reach the centre of  the host system and
to     cause     significant      changes     in     its     structure
\citep[VH08,][]{bt,kaza09}.

\begin{figure*}
\includegraphics[width=175mm,clip]{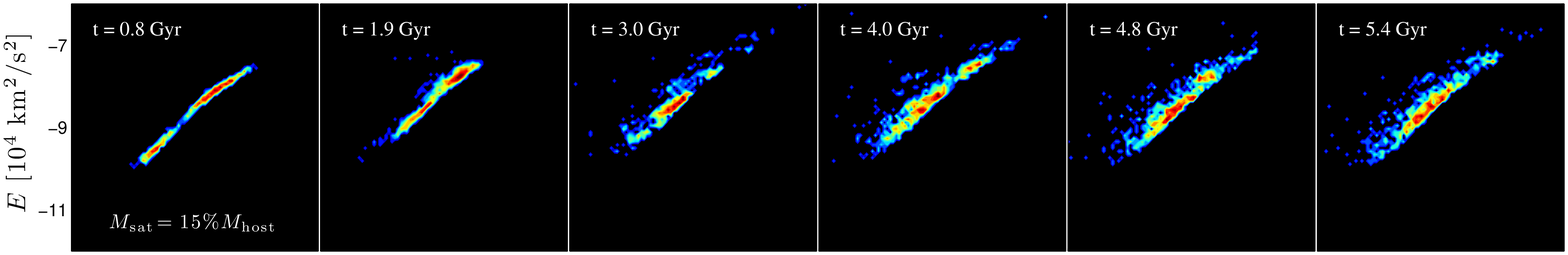}
\\
\includegraphics[width=175mm,clip]{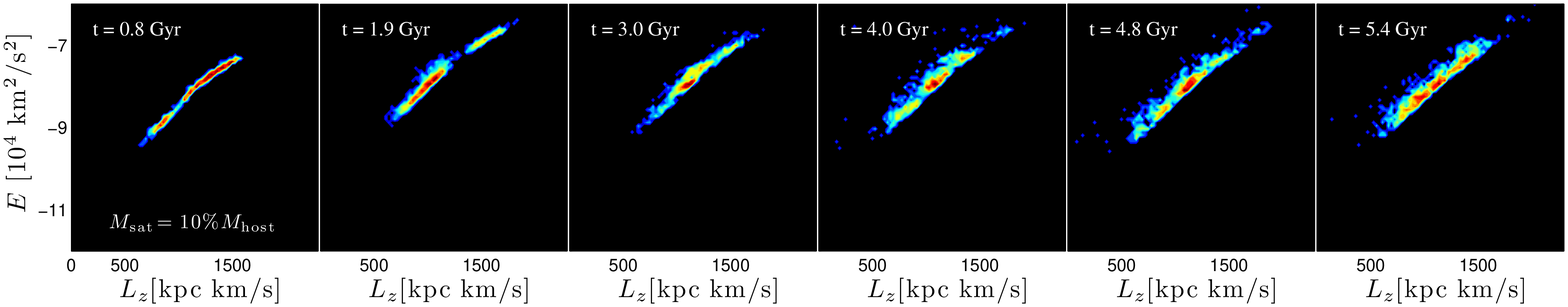}
\caption{Time  development  of  the  distribution  of  disc  particles
  located  within  a sphere  in  $E-L_{z}$  space,  obtained from  two
  $N$-body simulations  of mergers  with mass ratios  $f =  15\%$ (top
  panels) and $10\%$ (bottom panels).   The sphere is centred at 8 kpc
  from  the galactic  centre and  has a  radius of  2.5 kpc.   In both
  simulations, substructure  can be followed  for a total of  $\sim$ 4
  Gyr. The merger concludes at $t_{\rm merger} \approx 2.5$ and $4$ Gyr
  for the simulation with $f = 15\%$ and $f = 20\%$, respectively.}
\label{fig:diff_mass}
\end{figure*}

Figure~\ref{fig:diff_mass}  shows,  for  both  simulations,  the  time
evolution of the $E$-$L_{z}$  distribution of particles located inside
a sphere centred at 8 kpc from the galactic centre\footnote{Note that,
  as shown in  Figure 14 of VH08, the final  thick disc's scale length
  of the  $f=10\%$ simulation does not significantly  differs from the
  one  measured  on  the  $f=20\%$  simulation.}.   Both  spheres  are
rotating  with an  angular  frequency derived  from the  corresponding
velocity of their  LSR. Interestingly, even for the  simulation with a
mass ratio  $f =  10\%$ (bottom panels),  substructure can  be clearly
observed in the phase-space distribution of disc stars for a period of
about $4$ Gyr.  This is approximately  the same period of time that we
were able to follow substructure in our $f=20\%$ simulation.  This can
be explained as follows: Although  the strength of the perturbation on
the  host's disc  depends on  satellite's mass,  lower-mass satellites
take longer to completely sink onto the galactic centre.  For example,
in  our $f=10\%$  simulation $t_{\rm  merger}  \approx 4$  Gyr.  As  a
consequence, a lower mass satellite can excite the galactic disc for a
longer period of time \citep[see also][]{quill09}.  Note however that,
after $t_{\rm  merger}$, substructure can  be followed only  for $\sim
2.3$ and $\sim 0.9$ Gyr for the simulation with $f = 15\%$ and $10\%$,
respectively.   Low  mass  satellites  generates  density  waves  with
smaller amplitudes  and thus, their signatures  after $t_{\rm merger}$
disappear faster as they quickly get within the noise level.

\subsection{Contamination from satellite's debris}

In the analysis  carried out so far, only  particles from the galactic
disc were taken  into account. As described by  VH08, a large fraction
of  the  particles   from  the  satellite  ends  up   in  a  disc-like
configuration, with  the same spatial  orientation as the  final thick
disc.   Therefore, the  distribution of  the satellite's  particles in
$E$-$L_{z}$  space  could  partially  overlap with  the  corresponding
distribution of disc particles,  complicating the detection of density
waves.

The  distribution   of  debris  left   by  a  satellite  on   a  solar
neighbourhood-like  sphere, obtained from a simulation  similar to the
ones considered here, was already studied in detail by GH10.  The only
difference  between  their $f  =  20\%$ simulation and  ours  is  the
particle  resolution   of  the  satellite  galaxy.    To  enhance  the
signatures  of stellar  streams, the  satellite analysed  by  GH10 was
modelled with a number of particles ten times larger than the one used
here (see  also VH08).   Figure 13 of  GH10 shows the  distribution of
particles  from  both  systems,   in  four  different  projections  of
phase-space.  Not surprisingly, for a given energy, particles from the
satellites tend to  have smaller angular momentum than  those from the
disc (see  bottom middle panel). Even  in this simulation  with a much
higher  satellite's  resolution, substructure  from  the  disc can  be
easily identified in $E$-$L_{z}$ space. Note as well that substructure
in the disc can be clearly observed in apocentre vs.  pericentre space
(bottom left panel).

\section{Summary and Conclusions}
\label{sec:conclu}

In this  work we have explored  the impact that minor  mergers have on
the phase space  distribution of a galactic stellar  disc, such as the
one in the Milky Way.  Using numerical simulations, we showed that the
pseudo-integrals  of  motion   $E$-$L_{z}$  space  is  well-suited  to
identify density  waves in  the disc's velocity  field that  have been
generated by  the merger event.  In  this space, particles  in a given
volume  of physical  space are  distributed along  features  of nearly
constant  energy,   each  associated  with  a   different  wave.   The
separation between density waves  at different energy levels indicates
the  time since  the  disc  was perturbed  by  the merging  satellite.
Interestingly,  substructure  in  $E$-$L_{z}$  space  can  be  clearly
observed  even when  solar  neighborhood-like spheres  with radii  as
large as 2.5 kpc are considered.   This is in stark contrast with what
is observed in the $u$-$v$  plane, where the waves inside spheres with
radii larger than $\sim 0.25$ kpc interfere, wiping out all signatures
of ringing.  An important implication  of this result is that a larger
sample of  stars could be explored  to search for  signatures of minor
merger events in the Milky Way  thick disc.  This is relevant not only
in  the context  of the  forthcoming full  6D phase  space  {\it Gaia}
catalog, but  also for available  catalogs, such as those  provided by
RAVE \citep[see, e.g.,][]{maarten} and SDSS \citep{lee}.

By projecting the distribution  of particles located inside spheres of
2.5  kpc radius onto  $E$-$L_{z}$ space  we were  able, for  the first
time,  to study  the properties  of ringing  in  fully self-consistent
$N$-body simulations of disc  galaxies minor mergers.  We showed that,
in a simulation of a merger between a satellite galaxy and a host disc
with a  mass ratio $f=  20 \%$, signatures  of ringing can  be clearly
observed for a total of $\sim 4$ Gyr and for about $2.7$ Gyr after the
merger  has  concluded. It  is  remarkable  that  the $\lesssim  1000$
particles  enclosed in  our local  volumes  were enough  to resolve  a
wealth of substructure  in the resulting thick disc.   In addition, we
showed that  the in-fall trajectory  of the perturber  satellite gives
rise  to  an  azimuthal   dependence  of  the  structure  observed  in
$E$-$L_{z}$ space.   Note that we have only  explored distributions of
particles in this pseudo-integrals of motion space.  Information about
orbital  properties of  the  merging satellite,  such  as its  in-fall
trajectory or its impact parameter, could be obtained by exploring the
distributions  of   particles'  angular   phases  as  a   function  of
galactocentric azimuthal  angle. This will  be analyse in  a follow-up
study.

We have also explored the  impact that mergers of different total mass
ratios have  on the phase-space structure of  the disc. Interestingly,
mergers with  mass ratios  as small as  $f=10\%$ give rise  to density
waves in the velocity field of  the disc that can also be followed for
a total of  $\sim 4$ Gyr.  This is a consequence  of the longer period
of time  required by a  lower mass satellite  to fully merge  with its
host. On the other hand, once the merger concludes substructure can be
followed for only  $\sim 0.9$ and $\sim 2.3$  Gyr for simulations with
mass ratio of $f = 10$  and $15\%$, respectively. Note that these
  results  are  constrained by  the  low  particle  resolution of  our
  $N$-body simulations;  higher resolution simulations  could allow us
  to follow substructure for  longer periods of time. Thus, signatures
  of minor merger that have evolved for longer than 4 Gyr could
  be identified from the very  large and accurate sample of thick disc
  stars that  soon will  become available thanks  to {\it  Gaia}.

M09 showed  that a merger event  $\approx 1.9$ Gyr ago  could not only
explain the high  velocity streams seen in the  Solar Neighbourhood at
approximate    $v$   of    -60   \citep[HR    1614,][]{eggen92},   -80
\citep{ari_fu},  -100 (Arcturus),  and -160  km/s  \citep{klement} but
also  predicts four  new high  velocity streams  at $v  \approx$ −140,
-120,  40 and  60  km/s.  By  selecting  samples of  thick disc  stars
\citep{gc,sch06}  they  found  indications  of these  new  streams  in
velocity  space.    However,  their  results   are  not  statistically
significant since,  due to  the constraints on  the size of  the local
volume ($d_{\max}  = 0.08~\&~0.15$ kpc),  samples of only 451  and 766
stars were explored.  We are currently analysing the aforementioned 6D
phase space  catalogs from RAVE and  SDSS to search  for signatures of
these streams in $E$-$L_{z}$ space.

\section*{Acknowledgments}
We would like to thank Alice Quillen for  helpful comments.  FAG was
supported  through  the NSF  Office  of  Cyberinfrastructure by  grant
PHY-0941373,  and  by  the  Michigan State  University  Institute  for
Cyber-Enabled Research. BWO was supported in part by the Department of
Energy  through  the  Los  Alamos National  Laboratory  Institute  for
Geophysics and Planetary Physics.

\label{lastpage}

\begin{thebibliography}{aa}

\bibitem[Antoja et al.(2009)]{ant09}Antoja T., Valenzuela O., Pichardo B., Moreno E., Figueras F., Fern\'andez D., 2009, ApJ, 700, L78
\bibitem[Arifyanto \& Fuchs(2006)]{ari_fu} Arifyanto M.~I., Fuchs B., 2006, A\&A, 449, 533

\bibitem[Belokurov et al.(2006)]{belu06} Belokurov V. et al., 2006, ApJ, 642, L137
\bibitem[Benson et al.(2004)]{ben04} Benson A. J., Lacey C. G., Frenk  C. S., Baugh C. M., Cole S., 2004, MNRAS, 351, 1215
\bibitem[Benson(2005)]{ben05} Benson A. J., 2005, MNRAS, 358, 551
\bibitem[Binney \& Tremaine(2008)]{bt} Binney J., Tremaine S., 2008, Galactic Dynamics. Princeton Univ. Press, Princeton, NJ
\bibitem[Breddels et al.(2010)]{maarten} Breddels M. A., et al., 2010, A\&A, 511, A90
\bibitem[Brooks et al.(2011)]{brooks} Brooks, A., et al.. 2011, ApJ, 728, 51
\bibitem[Bullock \& Johnston(2005)]{bj05} Bullock J.~S., Johnston K.~V., 2005, ApJ, 635, 931

%
\bibitem[Carpintero \& Aguilar(1998)]{daniel} Carpintero D.~D., Aguilar L.~A., 1998, MNRAS, 298, 1
\bibitem[Cooper et al.(2010)]{cooper} Cooper A.~P. et al., 2010, MNRAS, 406,  744
\bibitem[Chapman et al.(2008)]{chap} Chapman S.~C. et al., 2008, MNRAS, 390, 1437

\bibitem[De Lucia \& Helmi(2008)]{lh} De Lucia G., Helmi A., 2008,  MNRAS, 391, 14
\bibitem[Dehnen(2000)]{walter} Dehnen W., 2000, AJ, 119, 800
\bibitem[Di Matteo et al.(2011)]{dm11} Di Matteo P., Lehnert M. D., Qu Y., van Driel W., 2011, A\&A, 525, L3

\bibitem[Eggen(1996)]{eggen} Eggen O.J., 1996, AJ, 112, 1595
\bibitem[Eggen(1992)]{eggen92} Eggen O.~ J., 1992, AJ, 104, 1906
\bibitem[Font et al.(2006)]{font} Font A.~S., Johnston K.~V., Bullock J.~S, Robertson B.~E., 2006, ApJ, 646, 886
\bibitem[Fux(2001)]{fux01} Fux, R., 2001, A\&A, 373, 511
\bibitem[Freeman \& Bland-Hawthorn(2002)]{FBH} Freeman K., Bland-Hawthorn J., 2002, ARA\&A, 40, 487

\bibitem[G\'{o}mez \& Helmi(2010)]{gh10} G\'{o}mez F.~A., Helmi A., 2010, MNRAS, 401, 2285
\bibitem[G\'{o}mez et al.(2010)]{gea10} G\'{o}mez F.~A., Helmi A., Brown A.~G.~A., Li Y.~S., 2010, MNRAS, 408, 935

\bibitem[Helmi \& de Zeeuw(2000)]{hz00} Helmi A., de Zeeuw P.~T., 2000, MNRAS, 319, 657
\bibitem[Helmi \& White(1999)]{hw} Helmi A., White S.~D.~M., 1999, MNRAS, 307, 495
\bibitem[Helmi et al.(1999)]{hwzz99} Helmi A., White S.~D.~M., de Zeeuw P.~T., Zhao H., 1999, Nature, 402, 53
\bibitem[Helmi et al.(2006)]{h06} Helmi A., Navarro J.~F., Nordstr\"om B., Holmberg J., Abadi M.~G.,  Steinmetz M., 2006, MNRAS, 365, 1309
\bibitem[Hernquist(1990)]{hernq} Hernquist L., 1990, ApJ, 356, 359

\bibitem[Ibata et al.(2003)]{ibata03} Ibata R., Irwin M.~J., Lewis G.~F., Ferguson A.~M.~N., Tanvir N., 2003, MNRAS, 340, L21
\bibitem[Ibata et al.(2001)]{ibata01a}Ibata R.~A., Irwin M.~J., Lewis G.~F., Ferguson A.~M.~N., Tanvir N., 2001a, Nature, 412, 49
\bibitem[Ibata, Gilmore \& Irwin(1994)]{ibata94} Ibata R.~A., Gilmore G., Irwin M.~J., 1994, Nature, 370, 194

\bibitem[Kazantzidis et al.(2009)]{kaza09} Kazantzidis S., Zentner A. R., Kravtsov A. V., Bullock J. S., Debattista V. P., 2009, ApJ, 700, 1896
\bibitem[Klement, Fuchs \& Rix(2008)]{klement}Klement R., Fuchs B., Rix H.-W., 2008, ApJ, 685, 261
\bibitem[Klement et al.(2009)]{klement09} Klement R. et al., 2009, ApJ, 698, 865

\bibitem[Lee et al.(2011)]{lee} Lee Y. S., et al., 2011, ApJ, 738, 187
\bibitem[Mart\'inez-Delgado et al.(2009)]{Mart09} Mart\'inez-Delgado D., Pohlen M., Gabany R.J., Majewski S.R., Pe\~narrubia, J., Palma C., 2009, ApJ, 692, 955
\bibitem[Mart\'inez-Delgado et al.(2008)]{Mart08} Mart\'inez-Delgado D., Pe\~narrubia J., Gabany R. J., Trujillo I., Majewski S. R., Pohlen M., 2008, ApJ, 689, 184
\bibitem[McConnachie et al.(2009)]{McCo09} McConnachie A.~W. et al., 2009, Nature, 461, 6673

\bibitem[Michel-Dansac et al.(2011)]{md11} Michel-Dansac L., Abadi M. G., Navarro J. F., Steinmetz M., 2011, MNRAS, 414, L1
\bibitem[Minchev et al.(2010)]{min10} Minchev I., Boily C., Siebert A., \& Bienayme O., 2010, MNRAS, 407, 2122
\bibitem[Minchev et al.(2009)]{min09} Minchev I., Quillen A. C., Williams M., Freeman K. C., Nordhaus J., Siebert A., Bienayme O., 2009, MNRAS, 396, L56
\bibitem[Minchev \& Quillen(2008)]{min08} Minchev I., Quillen A.~C., 2008, MNRAS, 386, 157
\bibitem[Mo, Mao \& White(1998)]{mo} Mo H. J., Mao S., White S. D. M., 1998, MNRAS, 295, 319

\bibitem[Navarro, Frenk \& White(1996)]{nfw} Navarro J.~F., Frenk C.~S., White S.~D.~M., 1996, ApJ, 462, 563
\bibitem[Navarro, Helmi \& Freeman(2004)]{nhf} Navarro J.F., Helmi A., Freeman K.~C., 2004, ApJL, 601, L43
\bibitem[Newberg et al.(2002)]{new02} Newberg H.~J. et al., 2002, ApJ, 569, 245
\bibitem[Nordstr\"{o}m et al.(2004)]{gc} Nordstr\"{o}m B., Mayor M., Andersen J., Holmberg  J., Pont F., J{\o}rgensen B. R., Olsen E. H., Udry S., Mowlavi N.: 2004, 2004, A\&A, 418, 989 
\bibitem[Noordermeer, Merrifield \& Arag{\'o}n-Salamanca(2008)]{nma} Noordermeer E., Merrifield M.~R., Arag{\'o}n-Salamanca, A. 2008, MNRAS, 388, 1381


\bibitem[Oh et al.(2008)]{oh} Oh S. H., Kim W.-T., Lee H. M., Kim J., 2008, ApJ, 683, 94

\bibitem[Perryman et al.(2001)]{gaia} Perryman M.~A.~C. et al., 2001, A\&A, 369, 339

\bibitem[Quillen et al.(2009)]{quill09} Quillen A. C., Minchev I.,  Bland-Hawthorn J., Haywood M., 2009, MNRAS, 397, 1599
\bibitem[Quillen et al.(2011)]{quill11} Quillen A. C., Dougherty J.,  Bagley M. B., Minchev I., Comparetta J., 2011, MNRAS, 417, 762
\bibitem[Quinn, Hernquist \& Fullagar(1993)]{qh93} Quinn P. J., Hernquist L., Fullagar D. P., 1993, ApJ, 403, 74

\bibitem[Sales  et al.(2008)]{sales08} Sales L.~V., et al. 2008, MNRAS, 389, 1391
\bibitem[Schuster et. al.(2006)]{sch06} Schuster W. J., Moitinho A., M{\'a}rquez A., Parrao L., Covarrubias E., 2006, A\&A 445, 939
\bibitem[Somerville et al.(2008)]{somer} Somerville R. S., et al., 2008, ApJ, 672, 776
\bibitem[Starkenburg et al.(2009)]{else09} Starkenburg E. et al., 2009, ApJ, 698, 567

\bibitem[Tutukov \& Fedorova(2006)]{tutu} Tutukov A. V., Fedorova A. V., 2006, Astron. Rep., 50, 785
\bibitem[Yanny et al.(2003)]{yanny03} Yanny B. et al., 2003, ApJ, 588, 824

\bibitem[Villalobos(2009)]{alvphd} Villalobos A., 2009, Ph.D. thesis, University of Groningen
\bibitem[Villalobos \& Helmi(2009)]{vh09} Villalobos A., Helmi A., 2009, MNRAS, 399, 166
\bibitem[Villalobos  \& Helmi(2008)]{vh08}  Villalobos  A., Helmi  A., 2008, MNRAS, 391, 1806


\bibitem[Williams et al.(2011)]{will11} Williams M.~E.~K, et al., 2011, ApJ, 728, 102
\bibitem[Williams et al.(2009)]{will09} Williams M. E. K., Freeman K. C., Helmi A., the RAVE collaboration, 2009, in IAU Symp. 254, The Galaxy Disk in Cosmological Context, ed. J. Andersen, J. Bland-Hawthorn, \& B. Nordstr{\"o}m, (Cambridge: Cambridge Univ. Press), 139

      

\end{thebibliography}
\end{document}